\documentclass[10pt]{article}

\usepackage[a4paper]{geometry}

\usepackage{makecell}
\usepackage{graphicx}
\usepackage{listings}
\usepackage{xcolor}
\usepackage{hyperref}
\usepackage{multicol}
\usepackage[backend=bibtex, style=nature]{biblatex}
\let\cite=\supercite

\title{Generating CodeMeta through declarative mapping rules: An open-ended approach using ShExML}
\author{Herminio García-González \\ \url{herminio.garciagonzalez@kazernedossin.eu}}
\date{\textit{Kazerne Dossin, Mechelen, Belgium}}

\begin{document}
\maketitle

\begin{abstract}
Nowadays, software is one of the cornerstones when conducting research in several scientific fields which employ computer-based methodologies to answer new research questions. However, for these experiments to be completely reproducible, research software should comply with the FAIR principles, yet its metadata can be represented following different data models and spread across different locations. In order to bring some cohesion to the field, CodeMeta was proposed as a vocabulary to represent research software metadata in a unified and standardised manner. While existing tools can help users to generate CodeMeta files for some specific use cases, they fall short on flexibility and adaptability. Hence, in this work, I propose the use of declarative mapping rules to generate CodeMeta files, illustrated through the implementation of three crosswalks in ShExML which are then expanded and merged to cover the generation of CodeMeta files for two existing research software artefacts. Moreover, the outputs are validated using SHACL and ShEx and the whole generation workflow is automated requiring minimal user intervention upon a new version release. This work can, therefore, be used as an example upon which other developers can include a CodeMeta generation workflow in their repositories, facilitating the adoption of CodeMeta and, ultimately, increasing research software FAIRness.

\end{abstract}

\lstdefinestyle{RDF}{
  classoffset=3,
  keywords={@prefix},
  keywordstyle=\color{gray},
  classoffset=1,
  morekeywords={:, ;, \,, .},
  classoffset=2,   
  keywordstyle=\color{darkgray}\bfseries,
  morekeywords={schema, dbr, xsd, a, rdf, rdfs, foaf, ex, sh},
  keywordstyle=\color{forestgreen},
  classoffset=4,
  identifierstyle=\color{black},
  sensitive=false,
  commentstyle=\color{forestgreen}\ttfamily,
  stringstyle=\color{brown}\ttfamily,
  breaklines=true,
  postbreak=\raisebox{0ex}[0ex][0ex]{\ensuremath{\color{gray}\hookrightarrow\space}},
  morestring=[b]',
  morestring=[b]",
  comment=[l]{\#\#},
  stringstyle=\color{Brown},
  identifierstyle=\color{darkblue},
  keywordstyle=\color{blush}\bfseries
}

\lstdefinestyle{RDFinline}{
  classoffset=3,
  keywords={@prefix},
  keywordstyle=\color{gray},
  classoffset=1,
  morekeywords={:, ;, \,, .},
  classoffset=2,   
  keywordstyle=\color{darkgray}\bfseries,
  morekeywords={schema, dbr, xsd, a, rdf, rdfs, foaf, ex, sh},
  keywordstyle=\color{forestgreen},
  classoffset=4,
  identifierstyle=\color{black},
  sensitive=false,
  commentstyle=\color{forestgreen}\ttfamily,
  stringstyle=\color{brown}\ttfamily,
  morestring=[b]',
  morestring=[b]",
  comment=[l]{\#\#},
  stringstyle=\color{Brown},
  identifierstyle=\color{darkblue},
  keywordstyle=\color{blush}\bfseries
}

\lstdefinestyle{ShExML}{
  classoffset=0,
  	keywords={PREFIX, SOURCE, ITERATOR, EXPRESSION, UNION, JOIN, FIELD, 
	MATCHER, AS, MATCHING, FUNCTIONS},
  	keywordstyle=\color{amaranth},
  classoffset=1,
  	morekeywords={<, >, [, ], ;, xpath:, jsonpath:, scala:, +},
  	keywordstyle=\color{darkgray}\bfseries,
  classoffset=0,
  	identifierstyle=\color{black},
  	sensitive=false,
  	commentstyle=\color{forestgreen}\ttfamily,
% 	commentstyle=\color{purple}\ttfamily,
  	stringstyle=\color{brown}\ttfamily,
  	showstringspaces=false,
  	breaklines=true,
  	postbreak=\raisebox{0ex}[0ex][0ex]{\ensuremath{\color{gray}\hookrightarrow\space}},
  	morestring=[b]',
  	morestring=[b]",
  	alsoletter={:},
  	comment=[l]{\#\#},
  	stringstyle=\color{Brown},
  	identifierstyle=\color{darkblue},
  	keywordstyle=\color{blush}\bfseries
}

\lstdefinestyle{ShExMLinline}{
  classoffset=0,
  	keywords={PREFIX, SOURCE, ITERATOR, EXPRESSION, UNION, JOIN, FIELD, 
	MATCHER, AS, MATCHING, FUNCTIONS},
  	keywordstyle=\color{amaranth},
  classoffset=1,
  	morekeywords={<, >, [, ], ;, xpath:, jsonpath:, scala:, +},
  	keywordstyle=\color{darkgray}\bfseries,
  classoffset=0,
  	identifierstyle=\color{black},
  	sensitive=false,
  	commentstyle=\color{forestgreen}\ttfamily,
% 	commentstyle=\color{purple}\ttfamily,
  	stringstyle=\color{brown}\ttfamily,
  	showstringspaces=false,
  	morestring=[b]',
  	morestring=[b]",
  	alsoletter={:},
  	comment=[l]{\#\#},
  	stringstyle=\color{Brown},
  	identifierstyle=\color{darkblue},
  	keywordstyle=\color{blush}\bfseries
}

\lstdefinestyle{bash}{
   language=Bash,
   showspaces=false,
   showstringspaces=false,
   basicstyle=\ttfamily,
   columns=flexible,
   stringstyle=\color{Brown},
   breaklines=true,
   postbreak=\raisebox{0ex}[0ex][0ex]{\ensuremath{\color{gray}\hookrightarrow\space}},
   keywordstyle=\color{darkblue}\ttfamily\textbf,
   commentstyle=\color{forestgreen}\ttfamily\textit
 }

\lstdefinelanguage{json}{
    frame=single,
    rulecolor=\color{black},
    showspaces=false,
    columns=flexible,
    showtabs=false,
    breaklines=true,
    postbreak=\raisebox{0ex}[0ex][0ex]{\ensuremath{\color{gray}\hookrightarrow\space}},
    breakatwhitespace=true,
    upquote=false,
    morestring=[b]",
    showstringspaces=false,
    stringstyle=\color{string},
    literate=
     *{0}{{{\color{numb}0}}}{1}
      {1}{{{\color{numb}1}}}{1}
      {2}{{{\color{numb}2}}}{1}
      {3}{{{\color{numb}3}}}{1}
      {4}{{{\color{numb}4}}}{1}
      {5}{{{\color{numb}5}}}{1}
      {6}{{{\color{numb}6}}}{1}
      {7}{{{\color{numb}7}}}{1}
      {8}{{{\color{numb}8}}}{1}
      {9}{{{\color{numb}9}}}{1}
      {\{}{{{\color{delim}{\{}}}}{1}
      {\}}{{{\color{delim}{\}}}}}{1}
      {[}{{{\color{delim}{[}}}}{1}
      {]}{{{\color{delim}{]}}}}{1},
}

\lstdefinestyle{jsoninline}{
    frame=single,
    rulecolor=\color{string},
    showspaces=false,
    showtabs=false,
    breaklines=true,
    postbreak=\raisebox{0ex}[0ex][0ex]{\ensuremath{\color{gray}\hookrightarrow\space}},
    breakatwhitespace=true,
    upquote=true,
    morestring=[b]",
    alsoletter={:},
    showstringspaces=false,
    stringstyle=\color{string},
    identifierstyle=\color{string},
    literate=
     *{0}{{{\color{numb}0}}}{1}
      {1}{{{\color{numb}1}}}{1}
      {2}{{{\color{numb}2}}}{1}
      {3}{{{\color{numb}3}}}{1}
      {4}{{{\color{numb}4}}}{1}
      {5}{{{\color{numb}5}}}{1}
      {6}{{{\color{numb}6}}}{1}
      {7}{{{\color{numb}7}}}{1}
      {8}{{{\color{numb}8}}}{1}
      {9}{{{\color{numb}9}}}{1}
      {\{}{{{\color{delim}{\{}}}}{1}
      {\}}{{{\color{delim}{\}}}}}{1}
      {[}{{{\color{delim}{[}}}}{1}
      {]}{{{\color{delim}{]}}}}{1},
}

\newcommand\YAMLcolonstyle{\color{black}\mdseries}
\newcommand\YAMLkeystyle{\color{black}\bfseries}
\newcommand\YAMLvaluestyle{\color{darkblue}\mdseries}

\makeatletter

% here is a macro expanding to the name of the language
% (handy if you decide to change it further down the road)
\newcommand\language@yaml{yaml}

\expandafter\expandafter\expandafter\lstdefinelanguage
\expandafter{\language@yaml}
{
  keywords={true,false,null,y,n},
  keywordstyle=\color{darkgray}\bfseries,
  basicstyle=\YAMLkeystyle,                                 % assuming a key comes first
  sensitive=false,
  comment=[l]{\#},
  morecomment=[s]{/*}{*/},
  commentstyle=\color{purple}\ttfamily,
  stringstyle=\YAMLvaluestyle\ttfamily,
  showstringspaces=false,
  breaklines=true,
  postbreak=\raisebox{0ex}[0ex][0ex]{\ensuremath{\color{gray}\hookrightarrow\space}},
  moredelim=[l][\color{orange}]{\&},
  moredelim=[l][\color{magenta}]{*},
  moredelim=**[il][\YAMLcolonstyle{:}\YAMLvaluestyle]{:},   % switch to value style at :
  morestring=[b]',
  morestring=[b]",
  identifierstyle=\color{Brown}\bfseries,
  literate =    {---}{{\ProcessThreeDashes}}3
                {>}{{\textcolor{red}\textgreater}}1     
                {|}{{\textcolor{red}\textbar}}1 
                {\ -\ }{{\mdseries\ -\ }}3,
}

% switch to key style at EOL
\lst@AddToHook{EveryLine}{\ifx\lst@language\language@yaml\YAMLkeystyle\fi}
\makeatother

\newcommand\ProcessThreeDashes{\llap{\color{cyan}\mdseries-{-}-}}

\definecolor{delim}{RGB}{20,105,176}
\definecolor{numb}{RGB}{106, 109, 32}
\definecolor{string}{rgb}{0.64,0.08,0.08}

\definecolor{forestgreen}{RGB}{34,139,34}
\definecolor{orangered}{RGB}{239,134,64}
\definecolor{darkblue}{rgb}{0.0,0.0,0.6}
\definecolor{gray}{rgb}{0.4,0.4,0.4}
\definecolor{MidnightBlue}{rgb}{0.1, 0.1, 0.44}
\definecolor{Brown}{rgb}{0.59, 0.29, 0.0}
\definecolor{alizarin}{rgb}{0.82, 0.1, 0.26}
\definecolor{awesome}{rgb}{1.0, 0.13, 0.32}
\definecolor{amaranth}{rgb}{0.9, 0.17, 0.31}
\definecolor{bluegray}{rgb}{0.4, 0.6, 0.8}
\definecolor{blush}{rgb}{0.87, 0.36, 0.51}
\definecolor{darkgray}{rgb}{0.66, 0.66, 0.66}
\definecolor{pblue}{rgb}{0.13,0.13,1}
\definecolor{pgreen}{rgb}{0,0.5,0}
\definecolor{pred}{rgb}{0.9,0,0}
\definecolor{pgrey}{rgb}{0.46,0.45,0.48}
\definecolor{charcoal}{rgb}{0.21, 0.27, 0.31}

\section{Introduction}\label{sec:introduction}
When conducting research, an increasing number of fields are relying on computer software to solve – or prove – their research questions \cite{Hannay09}. Many times this software is self-produced as part of the research activities for which is commonly referred as research software \cite{Gruenpeter2021}. One of the key aspects when promoting Open Science is to ensure the general availability of the produced software so that the facts derived from it can be reproduced and contrasted \cite{Wilkinson17}. However, in order to fulfil this realisation research software must be Findable, Accessible, Interoperable and Reusable (FAIR) \cite{Wilkinson2016,Lamprecht20}.

Making research software more FAIR involves many different tasks and good practices which are in many occasions difficult to organise and homogenise across domains \cite{ChueHong2022}. In an effort to shed some light on this field, the FAIR-IMPACT project (\url{https://www.fair-impact.eu/}) undertook the creation of a set of best practices to improve the collection and curation of research software metadata which have been gathered in the Research Software MetaData (RSMD) guidelines \cite{Gruenpeter2024}. Nevertheless, as identified in \cite{Gruenpeter2024}, metadata for research software can be classified attending to its location: intrinsic (when it is packed with the source code) and extrinsic (when it is located externally to the source code).

As a result, many vocabularies and ontologies define different stages and parts of the software lifecycle but their interoperability is not guaranteed. In order to bring some cohesion to research software metadata definition, CodeMeta \cite{codemeta} proposed a crosswalk between different existing metadata formats to enhance their interoperability. This crosswalk would later derive in the CodeMeta schema, prescribing the use of the JavaScript Object Notation for Linked Data (JSON-LD) as the serialisation syntax. While some of the existing tools are able to generate CodeMeta files from different formats, they are mainly focused around one-to-one conversions, require the intervention of the user for some of the reconciliation process or lack automation or adaptability capabilities.

On the Semantic Web field \cite{berners2001semantic}, recent developments proposed the use of declarative mapping rules for the integration of heterogeneous data sources under a single Knowledge Graph (KG) given their enhanced readability, flexibility, maintainability, shareability and reusability over \textit{ad-hoc} approaches \cite{vanAssche2023} (making, incidentally, the whole data integration process more FAIR). Hence, in this paper, I propose the use of declarative mapping rules for the creation of valid CodeMeta files which is exemplified by the implementation of some of the existing CodeMeta crosswalks in the Shape Expressions Mapping Language (ShExML) \cite{Garcia-Gonzalez20}. Besides, the crosswalks are then expanded and merged to support the creation of CodeMeta files for two existing repositories maintained by the author of this paper. Upon these examples, practitioners could adapt them to their own needs given their increased adaptability and flexibility. Moreover, a GitHub Actions based workflow is presented which allows to automatically create a CodeMeta file, validate it using the Shapes Constraint Language (SHACL) \cite{shaclSpecification} and Shape Expressions (ShEx) \cite{Prudhommeaux14}, and upload it to the targeted repository, reducing the friction of maintaining the proposed solution. Thus, I see this endeavour as a first step to offer a CodeMeta generation workflow under the form of declarative mapping rules, that could facilitate the generation of CodeMeta files from existing heterogeneous extrinsic metadata providers while also having the ability of being fully automated.

The rest of this paper is structured as follows: Section \ref{sec:background} explores the background and the related work, in Section \ref{sec:methods} the followed method is described alongside additional resources used for its validation and automation, while Section \ref{sec:results} explores the results obtained after its implementation in two open-source repositories. Finally, Section \ref{sec:conclusions} draws the future lines of work and introduces the conclusions obtained from this work.

\section{Background}\label{sec:background}
In this section, I briefly introduce the CodeMeta schema, followed by an explanation of ShExML main concepts, allowing readers to fully understand the proposed method. Furthermore, the related work is discussed, highlighting how the current approach expands the state of the art.

\subsection{Brief introduction to CodeMeta}\label{sec:codemeta}
As introduced earlier, research software is widely used to support research in different scientific fields. However, its dissemination and preservation are far from being standardised which hinders its reusability and discoverability. Some recent advancements and best practices have slightly alleviated this problem in which academics deposit the associated resources in online repositories, such as: GitHub (\url{https://github.com/}), Zenodo (\url{https://zenodo.org/}) and FigShare (\url{https://figshare.com/}), amongst others. Inevitably, each platform uses its own set of metadata (mainly extrinsic metadata enriched by some intrinsic metadata) – and follows its own in-house vocabularies or domain models – making, consequently, the interoperability between different services a rather difficult task. When it comes to software creation, different technologies also use different means of representing intrinsic metadata. As an example, Maven-supported Java Virtual Machine (JVM) projects lean on the use of Maven's Project Object Model (POM) whereas npm-based JavaScript developments make use of the npm's package.json file. Due to the unlimited combinations of extrinsic and intrinsic metadata providers, ensuring the FAIRness of research software becomes a daunting endeavour.

In an effort to bring some coherence and consistency on the description of research software, CodeMeta was proposed as a standardised vocabulary able to mediate and conciliate between various metadata formats for research software. It leverages the \lstinline[style=RDFinline, frame=single, basicstyle=\normalsize\ttfamily]{schema:SoftwareSourceCode} and \lstinline[style=RDF, frame=single, basicstyle=\normalsize\ttfamily]{schema:SoftwareApplication} classes of the schema.org  vocabulary (\url{https://schema.org/}), expanding them with some specific attributes and advocating for the use of a JSON-LD syntax which can appeal to practitioners unfamiliar with Semantic Web technologies (see Listing \ref{lst:crosswalkGithubShExML} for an example).

\begin{lstlisting}[
 caption={Excerpt of an output following the CodeMeta vocabulary for the ShExML engine based on the crosswalk for GitHub defined in the CodeMeta webpage. ... denotes an omitted part of the file for the sake of brevity.}, label={lst:crosswalkGithubShExML}, language=json, frame=single,
  basicstyle=\scriptsize\ttfamily
]
{
  "id" : "http://example.org/ShExML",
  "type" : "SoftwareSourceCode",
  "author" : {
    "id" : "https://github.com/herminiogg",
    "type" : "Person",
    "name" : "herminiogg"
  },
  "codeRepository" : "https://github.com/herminiogg/ShExML",
  "dateCreated" : "2018-02-22",
  "dateModified" : "2025-07-18",
  "description" : "A heterogeneous data mapping language based on Shape Expressions",
  "downloadUrl" : "https://github.com/herminiogg/ShExML/releases",
  "identifier" : "122470958",
  "license" : "https://api.github.com/licenses/mit",
  "name" : "ShExML",
  "programmingLanguage" : "Scala",
  "issueTracker" : "https://github.com/herminiogg/ShExML/issues",
  "@context" : {
    ...
  }
}
\end{lstlisting}

One of the main features of CodeMeta is its possibility to serve as an interoperability layer between different metadata providers for which the per-platform crosswalks (\url{https://codemeta.github.io/crosswalk/}) for diverse metadata formats are one of the key aspects, ensuring that transformations to CodeMeta are performed in a systematic manner. In fact, CodeMeta was initially devised as an overarching crosswalk which takes into account all the existing properties from various standards and captures them in a unified vocabulary. This guarantees the fully coverage of the existing formats semantics while capturing their equivalences in other platforms.

As mentioned before, a set of tools is offered which alleviates the creation of CodeMeta files for different use cases and promotes its adoption. Moreover, this specification is being gradually integrated in some repositories, such as the Software Heritage Archive (\url{https://archive.softwareheritage.org/}) and HAL (\url{https://hal.science/}), or is planned to be integrated (like in the case of Zenodo).

\subsection{Brief introduction to ShExML}
ShExML is a language that allows to integrate heterogeneous data sources (such as XML, JSON, CSV and relational databases) into a single Resource Description Framework (RDF) file. For doing so, it leans on an easy-to-use syntax \cite{Garcia-Gonzalez20} inspired by ShEx, extending and adapting its syntax to the specific purpose of mapping data. Following this rationale, the language constructions are divided into two main categories, declarations and generators, which cover the two main concerns of data mapping activities, that is, extracting and generating data. 

The declarations part is broadly composed by: prefixes (letting users define shortcuts for the widely used Internationalised Resource Identifiers (IRIs) in the output), sources (defining the input data sources to be used to extract the actual data and where they are located), functions (which allow to extend the base functionality of the language), iterators (which define how the values will be extracted from the different input files) and expressions (allowing to merge iterators or results from different files). The iterators are, in turn, composed by other iterators (in order to facilitate a hierarchical lookup) and fields (which will tell the engine to extract a final value from the input file). Iterators and fields make use of existing query languages like JSONPath for consulting JSON files or XPath for XML files.

For defining the generation of an RDF graph, ShExML borrows the concept of shapes. Each shape will define how a subject will be generated with respect to their linked predicates and objects. Therefore, subjects and objects will indicate a prefix for the term generation (optional for objects, in which case the engine will create a literal) and a statement extracting the final values based on the expressions, iterators and fields defined in the declarations part. Predicates are predefined values following the Turtle syntax of prefix plus value. Similarly, it is also possible to use a predefined value in the form of a term or a literal as an object which the engine will include verbatim in the output. Moreover, shapes can be linked to indicate to the engine that an entity is linked to another one (using the subject-object RDF linkage based on IRIs).

A more profound explanation of the possibilities of ShExML and further features can be consulted on its specification \cite{shexml-spec} and an example of a ShExML input can be seen in Listing \ref{lst:crosswalkGithubShExMLMapping}.

\subsection{Related work}
Since CodeMeta inception, different tools were developed with the aim to facilitate its generation, and foster its adoption. Such tools can be categorised in three big groups (although they can pertain to more than one group at the same time): transformations from and to CodeMeta, management tools and auxiliary tools helping to draft a CodeMeta file.

\textbf{Transformations from and to CodeMeta:}
Bolognese \cite{bolognese} is a Ruby library that reads and writes DOI metadata in different formats, including CodeMeta. It uses crosscite as the internal model. Codemetar \cite{Boettiger2017, codemetar} allows to generate, parse and modify CodeMeta files automatically for R packages. Similarly, codemetapy \cite{codemetapy} converts from several metadata specifications (e.g., pip packages, Maven, NodeJS packages, etc.) to CodeMeta. The conversion of Citation File Format (CFF) files to other formats (including CodeMeta) is tackled by cffconvert \cite{cffconvert}. In order to centralise the authors' and contributors' information, tributors \cite{tributors} allows to maintain a single file for them and export it as CodeMeta (amongst other available formats). Extracting data from GitHub repositories and README files is the approach followed by the Software Metadata Extraction Framework (SoMEF) \cite{Mao19} which is then able to generate outputs following the Software Description Ontology \cite{software-description-ontology} or as CodeMeta. As a more overarching solution, codemeta-harvester \cite{codemeta-harvester} allows to retrieve various metadata formats and reconcile them in a single CodeMeta file by leveraging codemetapy, cffconvert and SoMEF (for extracting metadata from README files).

\textbf{Management tools:}
Both codemetar and codemetapy provide tools to work with CodeMeta files (including reading and manipulating existing ones) on top of the offered one-to-one transformations. Built on top of codemetapy and codemeta-harvester, codemeta-server \cite{codemeta-server} allows to construct a catalogue of tools based on CodeMeta. In addition, it provides a server and an in-memory triple store, allowing to perform queries using an Application Programming Interface (API) or a SPARQL endpoint. HERMES \cite{Kernchen2025} was born as a solution for publishing research software with rich metadata through Continuous Integration (CI) pipelines, it does so by pulling different metadata formats, pushing the merge results through a curation phase and automatically submitting the enriched research software to a publication repository (e.g., InvenioRDM (\url{https://inveniosoftware.org/products/rdm/})).

\textbf{Auxiliary tools:}
CodeMeta generator \cite{codemeta-generator} is a web-based application that generates CodeMeta files based on the users' input in a guided web form. As a complement, it is also able to validate the resulting file and import existing ones for their modification. Auto CodeMeta generator \cite{auto-codemeta-generator} was born as a fork of CodeMeta generator incorporating the possibility to automatically retrieve the data from a GitHub or GitLab repository, easing up the initial introduction of data. In a similar fashion, the Software Metadata Extraction and Curation Software (SMECS) \cite{ferenz2025} gathers different metadata formats (at the moment of writing, GitHub/GitLab, CFF and CodeMeta), presents them through a web-based user interface for further curation by the user and generates a final CodeMeta file. These three tools can also be considered as management tools or, in the case of Auto CodeMeta generator and SMECS, as transformation ones as well. However, they are included under this category as their primary goal is to help users to draft a valid CodeMeta file involving the less code possible and guiding them through the different sections.

Many of the presented tools only allow to perform a transformation from one metadata format to CodeMeta at a time or manage existing CodeMeta files or parts of them.  Only a handful of the presented tools allow to map and reconcile different formats in a single CodeMeta file, based on the use of a precedence-based reconciliation algorithm, meaning that higher-priority metadata providers will overwrite the attributes from less relevant ones, or in the involvement of a human through a curation phase. In contrast, the solution proposed in this paper enables a higher degree of flexibility by letting users map and merge different metadata sources in a single CodeMeta output with a very high level of granularity, yet in a more shareable and understandable format. Furthermore, using declarative mapping rules to define the generation of CodeMeta enables the inclusion of additional properties whenever they are not available in a provider, similar to what can be achieved with some of the presented management and auxiliary tools but with a more straightforward automation capability.

\section{Methods}\label{sec:methods}
In this section, the methodology followed to produce CodeMeta using ShExML is introduced \cite{codemeta-shexml}. Firstly, the implementation for three crosswalks is presented to continue with the development of mappings for two real projects (using and merging the previously created crosswalks mappings). Finally, the workflow will be enriched with: a JSON-LD framing process (to cater for the specific CodeMeta syntax), the validation of the generated outputs using SHACL and ShEx, and the final integration of the whole method in a GitHub Actions based workflow, making the final solution automated.

\subsection{Crosswalks: Maven, GitHub and Zenodo}
From the crosswalks available on the CodeMeta webpage, I have selected those that were accessible in the context of the ShExML engine project \cite{Garcia-Gonzalez25a, shexml-engine}, namely: GitHub, Maven and Zenodo. The implementations are done following a one-to-one translation into ShExML mapping rules, meaning that only the attributes present in the crosswalk are introduced in the mapping, even though some other attributes can also be included, or replaced by more meaningful ones depending on the context, as it will be further detailed in Section \ref{sec:generatingFullCodeMeta}. Table \ref{tab:crosswalksImplementationPerPlatform} summarises the established mappings between the three different metadata providers attributes and their counterparts in CodeMeta, based on the CodeMeta crosswalks and including how they overlap with each other.

\begin{lstlisting}[
 caption={ShExML mapping file used to convert the ShExML engine extrinsic metadata extracted from the GitHub API to CodeMeta.}, label={lst:crosswalkGithubShExMLMapping}, style=ShExML, frame=single,
  basicstyle=\scriptsize\ttfamily
]
PREFIX codemeta: <https://w3id.org/codemeta/3.0/>
PREFIX schema: <http://schema.org/>
PREFIX ex: <http://example.org/>
PREFIX xsd: <http://www.w3.org/2001/XMLSchema#>
PREFIX gh: <https://github.com/>
SOURCE repo_info <https://api.github.com/repos/herminiogg/ShExML>
FUNCTIONS helper <scala: ../functions.scala>
ITERATOR gh <jsonpath: $> {
    FIELD id <id>
    FIELD name <name>
    FIELD description <description>
    FIELD codeRepository <html_url>
    FIELD language <language>
    FIELD releases_url <releases_url>
    FIELD author <owner.login>
    FIELD dateCreated <created_at>
    FIELD dateModified <updated_at>
    FIELD keywords <topics>
    FIELD license <license.url>
    FIELD issueTracker <issues_url>
}
EXPRESSION md <repo_info.gh>

schema:SoftwareSourceCode ex:[md.name] {
    a schema:SoftwareSourceCode ;
    schema:identifier [md.id] ;
    schema:name [md.name] ;
    schema:description [md.description] ;
    schema:codeRepository [md.codeRepository] ;
    schema:programmingLanguage [md.language] ;
    schema:downloadUrl [helper.normalizeGitHubAPIUrl(md.releases_url)] ;
    schema:dateCreated [helper.onlyDate(md.dateCreated)] xsd:date ;
    schema:dateModified [helper.onlyDate(md.dateModified)] xsd:date ;
    schema:keywords [md.keywords] ;
    schema:license [md.license] ;
    codemeta:issueTracker [helper.normalizeGitHubAPIUrl(md.issueTracker)] ;

    schema:author @schema:Person ;
}

schema:Person gh:[md.author] {
    a schema:Person ;
    schema:name [md.author] ;
}

\end{lstlisting}

\textbf{GitHub}: This mapping uses the GitHub API, and its reponses in the JSON format, to transform the targeted attributes of the crosswalk (extracted through the ShExML iterator) to their counterpart CodeMeta attributes (through two linked shapes). The only special conversion relates to the use of a function to convert from the date format used by GitHub, following the ISO 8601 ``yyyy-MM-dd'T'HH:mm:ss'Z{'}'' format, to the CodeMeta ``yyyy-MM-dd'' date format. The resulting mapping file can be seen in Listing \ref{lst:crosswalkGithubShExMLMapping}.

\textbf{Maven}: The Maven mapping follows a similar approach to the GitHub one, however, in its case the POM file uses an XML format. Therefore, the extraction queries are performed using XPath and, due to the involvement of namespaces, the queries need to include some additional affordances to effectively retrieve the data. Besides, in this specific implementation there were no prerequisites in the POM file, hence the dependencies are mapped here instead as they can be more wide-applicable to other projects. As the \lstinline[style=jsoninline, frame=single, basicstyle=\normalsize\ttfamily]{softwareRequirements} term expects an entity of the type \lstinline[style=ShExML, frame=single, basicstyle=\normalsize\ttfamily]{schema:SoftwareSourceCode}, a new shape has to be created and linked to the main one. Ultimately, this results in the main \lstinline[style=ShExML, frame=single, basicstyle=\normalsize\ttfamily]{schema:SoftwareSourceCode} entity having various nested \lstinline[style=ShExML, frame=single, basicstyle=\normalsize\ttfamily]{schema:SoftwareSourceCode} entities listed as software requirements.

\textbf{Zenodo}: For converting from Zenodo to CodeMeta, the mapping follows the same principles described for the GitHub and Maven ones. The only particularity of this mapping relates to the nesting of three consecutive shapes to cover the \lstinline[style=jsoninline, frame=single, basicstyle=\normalsize\ttfamily]{author} term and then, once again, to capture the \lstinline[style=jsoninline, frame=single, basicstyle=\normalsize\ttfamily]{affiliation} of the said \lstinline[style=jsoninline, frame=single, basicstyle=\normalsize\ttfamily]{author}. In CodeMeta parlance, the \lstinline[style=ShExML, frame=single, basicstyle=\normalsize\ttfamily]{schema:SoftwareSourceCode} class is linked to a \lstinline[style=ShExML, frame=single, basicstyle=\normalsize\ttfamily]{schema:Person}, and the latter, on its turn, to a \lstinline[style=ShExML, frame=single, basicstyle=\normalsize\ttfamily]{schema:Organization}.

These crosswalks can be easily adapted to other projects by changing the input source to that of another project within the same metadata provider. While they are not fully comprehensive with respect to all the possible terms that can be linked towards the CodeMeta specification, as neither the crosswalks are, they offer a minimal and straightforward implementation that can be later expanded according to the specific use case. At the same time, they constitute the building blocks to elaborate more complex and unified ShExML mapping files able to generate CodeMeta outputs from various and heterogeneous metadata providers.

\begin{table}[]
\centering
\resizebox{\textwidth}{!}{%
\begin{tabular}{lllll}
\textbf{CodeMeta class}     & \textbf{CodeMeta attribute}  & \textbf{GitHub}          & \textbf{Zenodo}                                         & \textbf{Maven}                \\ \hline
SoftwareSourceCode & identifier          & id              & id                                             & groupId              \\
                   & name                & name            & title                                          & name                 \\
                   & description         & description     & metadata.description                           & description          \\
                   & codeRepository      & html\_url       & metadata.related\_identifiers.isSupplementTo   & scm/url              \\
                   & programmingLanguage & language        &                                                &                      \\
                   & downloadUrl         & releases\_url    & metadata.related\_identifiers.isIdenticalTo    &                      \\
                   & author              & owner.html\_url &                                                &                      \\
                   & dateCreated         & created\_at     &                                                &                      \\
                   & dateModified        & updated\_at     &                                                &                      \\
                   & keywords            & topics          & metadata.keywords                              &                      \\
                   & license             & license.url     & metadata.license.id                            & licenses/license/url \\
                   & issueTracker        & issues\_url     & metadata.related\_identifiers.isSupplementedBy & issueManagement      \\
                   & continuousIntegration &               &                                                & ciManagement         \\
                   & applicationCategory &                 & communities                                    &                      \\
                   & datePublished       &                 & metadata.publication\_date                     &                      \\
                   & softwareRequirements&                 &         										& (see SoftwareSourceCode (dependencies) class)                      \\
                   & funder              &                 & metadata.contributors.funder                   &                      \\
                   & version             &                 & metadata.version                               & version              \\
                   & funding             &                 & metadata.grants.title                          &                      \\
                   & author              & (see Person class) & (see Person class)                             &                      \\
 \hline
Person             & name                & owner.login     & metadata.creators.name                         &                      \\
				   & identifier          &                 & metadata.creators.orcid                        &                      \\ 
                   & affiliation         &                 & (see Organization class)                       &                      \\ \hline
Organization       & name                &                 & metadata.creators.affiliation                  &                      \\ \hline
SoftwareSourceCode (dependencies) & name &                 &                                                & \makecell[l]{[dependencies/dependency/groupId,\\dependencies/dependency/artifactId]}       \\ 
                   & version             &                 &                                                & dependencies/dependency/version       \\ \hline
\end{tabular}
}
\caption{This table collects the equivalences followed in the implementation of the ShExML rules based on the official crosswalks published on the CodeMeta webpage. The . and / symbols used for the attributes of the metadata providers (i.e., GitHub, Zenodo and Maven) denote the access to a child item in JSON and XML respectively.}
\label{tab:crosswalksImplementationPerPlatform}
\end{table}

\subsection{Generating CodeMeta for the ShExML engine}\label{sec:generatingFullCodeMeta}
In order to offer a full example implementation and illustrate how the three previously developed crosswalks can be merged into a single solution, a mapping file was created for the ShExML engine. This unified mapping file is based on the three previously introduced crosswalks mappings but adapted and expanded to cover the specificity of the targeted project. In this sense, when more than one crosswalk implements the same term (see Table \ref{tab:crosswalksImplementationPerPlatform}) the most relevant one is selected, avoiding a duplication of values. This decision is based upon the semantics of the attribute but also, in case of semantically equivalent attributes, in a practical choice as to which value is easier to maintain for the end user. For example, the \lstinline[style=ShExML, frame=single, basicstyle=\normalsize\ttfamily]{codemeta:description} term is implemented both in the Zenodo and GitHub crosswalks but, in the case of the ShExML engine, the description on Zenodo refers to the release notes (as they are automatically pushed after the creation of a new GitHub release). Therefore, the GitHub value is preserved as the most relevant one. Similarly, the version attribute can be pulled from the three providers but it demonstrates to be more practical to extract the value from Maven as it is the original source for this data (as intrinsic metadata). Some crosswalks are also expanded to cover additional terms like, for example, \lstinline[style=ShExML, frame=single, basicstyle=\normalsize\ttfamily]{schema:releaseNotes} which is available under the GitHub API Releases endpoint. Moreover, in this implementation I also considered the Zenodo-generated DOI to be a better – and more universal – candidate for the \lstinline[style=ShExML, frame=single, basicstyle=\normalsize\ttfamily]{schema:identifier} term.

Aside from the mapping of values, ShExML also offers the option to use \textit{hardcoded} values for those that are not present in the available metadata providers. While this can be seen as a bad practice, it allows developers to have more control over some values, complement those that cannot be retrieved from any platform and centralise their generation within the CodeMeta generation workflow. In other words, this feature offers a means to provide intrinsic and curated metadata centralised in the mapping file itself whenever this is suitable and required. In a later phase, these can also be pulled from another source, such as: a personal webpage or Wikidata, depending on the user's context and preferences. In this specific case, I used this option to have more control over the authors' and contributors' information or to complement some additional values (like the \lstinline[style=ShExML, frame=single, basicstyle=\normalsize\ttfamily]{schema:applicationCategory} or \lstinline[style=ShExMLinline, frame=single, basicstyle=\normalsize\ttfamily]{codemeta:referencePublication} terms). The final mapping file can be consulted in Listing \ref{lst:codemetaShExMLEngine} and it can directly produce a JSON-LD output. Additionally, Table \ref{tab:finalShExMLCrosswalk} offers an overview of the metadata providers for each of the used CodeMeta attributes in the proposed mapping.

\begin{table}[]
\centering
\resizebox{0.7\textwidth}{!}{%
\begin{tabular}{lll}
\textbf{CodeMeta class}     & \textbf{CodeMeta attribute}   & \textbf{Source}           \\ \hline
SoftwareSourceCode & identifier           & Zenodo           \\
                   & name                 & Maven            \\
                   & description          & GitHub           \\
                   & codeRepository       & Maven and GitHub \\
                   & programmingLanguage  & GitHub           \\
                   & downloadUrl          & GitHub           \\
                   & dateCreated          & GitHub           \\
                   & dateModified         & GitHub releases  \\
                   & keywords             & GitHub           \\
                   & license              & GitHub           \\
                   & issueTracker         & GitHub           \\
                   & applicationCategory  & User        \\
                   & funder               & Zenodo           \\
                   & version              & Maven            \\
                   & funding              & Zenodo           \\
                   & continousIntegration & Maven            \\
                   & runtimePlatform      & User        \\
                   & referencePublication & User        \\
				   & releaseNotes 		  & GitHub releases \\
                   & developmentStatus    & User        \\ 
                   & softwareRequirements & (see SofwareSourceCode (dependencies) class)        \\ 
                   & author    & (see Person class)        \\ 
                   & author    & (see Role class)        \\ 
                   & contributor    & (see Person class)        \\ \hline
Person             & givenName            & User        \\
                   & familyName           & User        \\
                   & affiliation          & (see Organization class)        \\
                   & email                & User        \\
                   & identifier           & User         \\ \hline
Organization       & name                 & User        \\ \hline
Role               & roleName             & User        \\ \hline
SoftwareSourceCode (dependencies) & name                 & Maven            \\
                   & version              & Maven           \\ \hline
\end{tabular}
}
\caption{This table shows the origin of the data for the mapping rules developed to generate CodeMeta files for the ShExML engine. User as a source means that the data is directly introduced by the final user in the mapping rules.}
\label{tab:finalShExMLCrosswalk}
\end{table}

\begin{lstlisting}[
 caption={ShExML mapping rules used to generate a CodeMeta file for the ShExML engine using metadata provided by GitHub, Maven and Zenodo.}, label={lst:codemetaShExMLEngine}, style=ShExML, frame=single,
  basicstyle=\scriptsize\ttfamily, literate=
  {á}{{\'a}}1
  {í}{{\'i}}1
]
PREFIX codemeta: <https://w3id.org/codemeta/3.0/>
PREFIX schema: <http://schema.org/>
PREFIX ex: <http://example.org/>
PREFIX gh: <https://github.com/>
PREFIX ghhgg: <https://github.com/herminiogg/>
PREFIX hgg: <https://herminiogarcia.com/#>
PREFIX kd: <https://kazernedossin.eu/>
PREFIX niod: <https://niod.knaw.nl/>
PREFIX niod_staff: <https://niod.knaw.nl/en/staff/>
PREFIX xsd: <http://www.w3.org/2001/XMLSchema#>
SOURCE gh_info <https://api.github.com/repos/herminiogg/ShExML>
SOURCE gh_releases <https://api.github.com/repos/herminiogg/ShExML/releases>
SOURCE maven_info <https://repo1.maven.org/maven2/com/herminiogarcia/shexml_3/0.6.0/shexml_3-0.6.0.pom>
SOURCE zenodo_record <https://zenodo.org/api/records/17092549>
FUNCTIONS helper <scala: https://raw.githubusercontent.com/herminiogg/codemeta-shexml/main/functions.scala>
ITERATOR gh <jsonpath: $> {
    FIELD description <description>
    FIELD codeRepository <html_url>
    FIELD language <language>
    FIELD downloadUrl <downloads_url>
    FIELD author <owner.html_url>
    FIELD dateCreated <created_at>
    FIELD dateModified <updated_at>
    FIELD keywords <topics>
    FIELD license <license.url>
    FIELD issueTracker <issues_url>
}
ITERATOR maven <xpath: /node()[local-name(.)='project']> {
    FIELD name <node()[local-name(.)='name']>
    FIELD version <node()[local-name(.)='version']>
    FIELD continuousIntegration <node()[local-name(.)='continuousIntegration']>
    FIELD codeRepository <node()[local-name(.)='scm']/node()[local-name(.)='url']>
    ITERATOR softwareRequirements <node()[local-name(.)='dependencies']/node()[local-name(.)='dependency']> {
        FIELD groupdId <node()[local-name(.)='groupId']>
        FIELD artifactId <node()[local-name(.)='artifactId']>
        FIELD version <node()[local-name(.)='version']>
    }
}
ITERATOR info <jsonpath: $> {
    FIELD doi <doi_url>
    FIELD funding <metadata.grants.internal_id>
    FIELD funder <metadata.grants.funder.name>
    FIELD identifier <id>
    FIELD codeRepository <metadata.custom.code:codeRepository>
}
ITERATOR releases <jsonpath: $[0]> {
    FIELD publicationDate <published_at>
    FIELD releaseNotes <body>
}
EXPRESSION md <gh_info.gh UNION maven_info.maven UNION zenodo_record.info>
EXPRESSION release <gh_releases.releases>

schema:SoftwareSourceCode gh:[helper.getLocalPartGithubRepo(md.codeRepository)] {
    a schema:SoftwareSourceCode ;
    schema:name [md.name] ;
    schema:description [md.description] ;
    schema:dateCreated [helper.onlyDate(md.dateCreated)] xsd:date ;
    schema:license [md.license] ;
    schema:identifier [md.doi] ;
    schema:applicationCategory "Computer Science" ;
    schema:keywords [md.keywords] ;
    schema:funding [md.funding] ;
    schema:funder [md.funder] ;
    schema:codeRepository [md.codeRepository] ;
    codemeta:continuousIntegration [md.continuousIntegration] ;
    codemeta:issueTracker [helper.removeParametersGithubAPI(md.issueTracker)] ;
    schema:programmingLanguage [md.language] ;
    schema:runtimePlatform "JVM" ;
    schema:softwareRequirements @ex:Dependency ;
    schema:version [md.version] ;
    schema:downloadUrl [md.downloadUrl] ;
    codemeta:referencePublication "https://doi.org/10.7717/peerj-cs.318" ;
    schema:developmentStatus "active" ;
    schema:author @ex:Author ;
    schema:author @ex:AuthorRole ;
    schema:contributor @ex:Contributor ;
} 

ex:Release ghhgg:ShExML {
    schema:dateModified [helper.onlyDate(release.publicationDate)] xsd:date ;
    schema:releaseNotes [release.releaseNotes] ;
}

ex:Dependency ex:[md.softwareRequirements.artifactId] {
    a schema:SoftwareSourceCode ;
    schema:name [helper.concatenateMavenGroupAndArtifactIds(md.softwareRequirements.groupdId, md.softwareRequirements.artifactId)] ;
    schema:version [md.softwareRequirements.version] ;
}

ex:Author hgg:me {
    a schema:Person ;
    schema:givenName "Herminio" ;
    schema:familyName "García González" ;
    schema:affiliation @ex:KD ;
    schema:email "herminio.garciagonzalez@kazernedossin.eu" ;
    schema:identifier "https://orcid.org/0000-0001-5590-4857" ;
}

ex:AuthorRole _:mainAuthor {
    a schema:Role ;
    schema:roleName "Main author" ;
}

ex:KD kd:en {
    a schema:Organization ;
    schema:name "Kazerne Dossin" ;
}

ex:Contributor niod_staff:mikebryant {
    a schema:Person ;
    schema:givenName "Mike" ;
    schema:familyName "Bryant" ;
    schema:affiliation @ex:NIOD ;
    schema:email "m.bryant@niod.knaw.nl" ;
    schema:identifier "https://orcid.org/0000-0003-0765-7390" ;
}

ex:NIOD niod:en {
    a schema:Organization ;
    schema:name "NIOD Institute for War, Holocaust and Genocide Studies" ;
}
\end{lstlisting}

\subsection{Framing the resulting JSON-LD}
The results yielded by ShExML – and for that matter by other declarative mapping rules – are mainly based on the RDF data model and their different serialisations (i.e., Turtle, N-Triples, JSON-LD, etc.) and, therefore, when using the JSON-LD serialisation this output will normally be an unframed JSON-LD. Unfortunately, as a usability affordance, CodeMeta advocates for the usage of a framed JSON-LD based on the publicly available CodeMeta context (\url{https://w3id.org/codemeta/3.0/}). Consequently, the derived tools assume this syntax and will not always work with the unframed counterpart, despite it being a valid RDF graph.

\begin{figure}[h]
	\centering
  	\caption{Diagram of the complete workflow followed to generate CodeMeta files with ShExML, including the post-transformation step to frame the JSON-LD output.}
	\label{fig:generalWorkflow}
	\includegraphics[width=\textwidth]{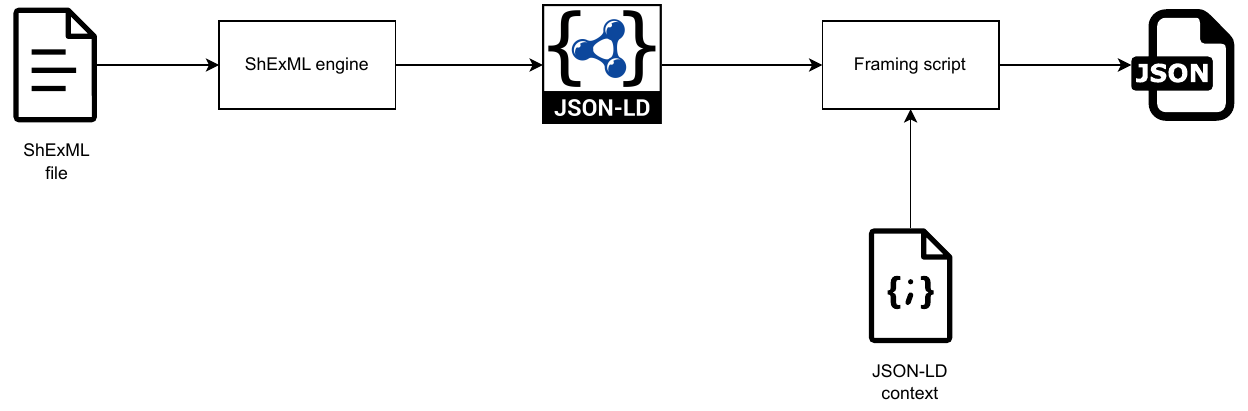}
\end{figure}

This is even more acute on account of the only existence of one validator integrated under the CodeMeta Generator web application which assumes this precept. Inevitably, this entails the creation of a framing JSON-LD algorithm which will be executed as a post-conversion process. Though this process could be included in the future in the ShExML engine, for the sake of its reproducibility and shareability (in similar contexts), it has been implemented as a Groovy script which takes two inputs: the unframed JSON-LD and a JSON-LD context. The latter is a JSON-LD file that defines how a user wants to shape the output in the framed JSON-LD version. When integrated in the general workflow (see Figure \ref{fig:generalWorkflow}) it allows to generate a compatible and valid CodeMeta file (see Listing \ref{lst:finalCodeMetaShExMLEngine}) that can be integrated in any online repository. This output is currently being used in the ShExML GitHub repository (see \url{https://github.com/herminiogg/ShExML/blob/master/codemeta.json}).

\begin{lstlisting}[
 caption={Excerpt of the final CodeMeta file generated for the ShExML engine after the execution of the mapping rules defined in Listing \ref{lst:codemetaShExMLEngine} and the post-transformation using the framing script. ... denotes an omitted part of the file for the sake of brevity.}, label={lst:finalCodeMetaShExMLEngine}, language=json, frame=single,
  basicstyle=\scriptsize\ttfamily, literate=
  {á}{{\'a}}1
  {í}{{\'i}}1
]
{
  "@context" : { ... },
  "id" : "https://github.com/herminiogg/ShExML",
  "type" : "SoftwareSourceCode",
  "applicationCategory" : "Computer Science",
  "author" : [ {
    "type" : "Role",
    "roleName" : "Main author"
  }, {
    "id" : "https://herminiogarcia.com/#me",
    "type" : "Person",
    "affiliation" : {
      "id" : "https://kazernedossin.eu/en",
      "type" : "Organization",
      "name" : "Kazerne Dossin"
    },
    "email" : "herminio.garciagonzalez@kazernedossin.eu",
    "familyName" : "García González",
    "givenName" : "Herminio",
    "identifier" : "https://orcid.org/0000-0001-5590-4857"
  } ],
  "codeRepository" : "https://github.com/herminiogg/ShExML",
  "contributor" : {
    "id" : "https://niod.knaw.nl/en/staff/mikebryant",
    "type" : "Person",
    "affiliation" : {
      "id" : "https://niod.knaw.nl/en",
      "type" : "Organization",
      "name" : "NIOD Institute for War, Holocaust and Genocide Studies"
    },
    "email" : "m.bryant@niod.knaw.nl",
    "familyName" : "Bryant",
    "givenName" : "Mike",
    "identifier" : "https://orcid.org/0000-0003-0765-7390"
  },
  "dateCreated" : "2018-02-22",
  "dateModified" : "2025-09-10",
  "description" : "A heterogeneous data mapping language based on Shape Expressions",
  "developmentStatus" : "active",
  "downloadUrl" : "https://api.github.com/repos/herminiogg/ShExML/downloads",
  "identifier" : "https://doi.org/10.5281/zenodo.17092549",
  "license" : "https://api.github.com/licenses/mit",
  "name" : "ShExML",
  "programmingLanguage" : "Scala",
  "releaseNotes" : "## What's Changed\r\n- Added a parellelisation option in the RDF conversion. You can decide which parts of the execution you want to run in parallel and the number of threads to be used (or let the engine decide based on you hardware specs).\r\n- Stdin can be used as input for the mapping rules or as a input source.\r\n- Some minor fixes and stability improvements.\r\n\r\n**Full Changelog**: https://github.com/herminiogg/ShExML/compare/v0.5.4...v0.6.0",
  "runtimePlatform" : "JVM",
  "softwareRequirements" : [ ... ],
  "version" : "0.6.0",
  "issueTracker" : "https://api.github.com/repos/herminiogg/ShExML/issues",
  "referencePublication" : "https://doi.org/10.7717/peerj-cs.318"
}
\end{lstlisting}

\subsection{Adapting to other projects: The case of DMAOG}
In order to demonstrate the adaptability of this approach, I modified the developed ShExML mapping rules for the ShExML engine (presented in Listing \ref{lst:codemetaShExMLEngine}) to accommodate the generation of a CodeMeta file for the DMAOG library \cite{dmaog}. Given that both projects are developed using the same technologies (i.e., written in Scala and managed with SBT) and published on the same platforms (i.e., using Maven Central, GitHub and Zenodo), the required adaptations are minimal. Namely, only 6 lines of code required modification (4 of them pertaining to the calls to the APIs), plus the removal of the additional contributor in ShExML (not present in DMAOG), to effectively generate a valid CodeMeta file for DMAOG (see the generated file at \url{https://github.com/herminiogg/dmaog/blob/main/codemeta.json}).

It is worth noting that adapting this conversion to other projects, even when using the same – or similar – technologies, will require more modifications. Nevertheless, this demonstrates that once adapted to a specific development methodology and team, more projects can be added with a minimal effort and time investment. Moreover, given the enhanced flexibility and adaptability of declarative mapping rules, this method can be effectively used and extended to cover other arbitrary projects. As such, the examples drafted here can be used as a blueprint for supporting the integration of CodeMeta on more projects and increase the adoption of CodeMeta at scale.

\subsection{Validating CodeMeta files}
In order to ensure that the generated CodeMeta files meet the requirements imposed by the vocabulary, the designed workflow needs to verify that each new generation keeps its valid form, which can become invalid either by a small change in the mapping rules or in the provided data. Some of the available tools used to offer a validation option, however, presently only the CodeMeta Generator tool still offers this possibility. Unfortunately, the validation algorithm is enclosed within the webpage code which hampers a straightforward reusability by other solutions. To this end, I have developed a set of validation rules in SHACL and ShEx, making the process more reusable (as it is based on wide-used technologies in the Semantic Web community, being SHACL a W3C recommendation), more adaptable for future versions of CodeMeta, and more widely applicable (due to the plethora of SHACL and ShEx implementations covering different programming languages and technical infrastructures). Both schemas cover the full specification provided by the CodeMeta consortium and have been tested over the outputs of all the mappings described in this paper. 

This additional validation step is provided notwithstanding the already existing validator and should serve as a more direct, actionable and productive approach than validating the output on the CodeMeta Generator website upon each generation. Nevertheless, for maximum compatibility, users are still encouraged to run a final validation on the mentioned website as it has been done with all the JSON-LD framed outputs presented in this paper and in its companion repository.

\subsection{Automating the CodeMeta generation workflow}
Even though the introduced workflow can be operated manually, its true potential comes from the possibility to automate it. For this purpose, the whole process is enclosed in a Bash script (see Listing \ref{lst:scriptGenerationCodeMeta}) which contains the execution of the ShExML engine using the developed mapping rules as input. Its output is then transformed using the aforementioned Groovy script to comply with the preceptive framed JSON-LD syntax of CodeMeta. Additionally, the script checks whether the ShExML engine library exists within the working directory, and downloads it otherwise. This encapsulation streamlines the generation of CodeMeta files for research software whenever an updated version is required.

\begin{lstlisting}[
 caption={Bash script used to automatically run the described workflow to generate CodeMeta files for the ShExML engine.}, label={lst:scriptGenerationCodeMeta}, style=bash, frame=single,
  basicstyle=\scriptsize\ttfamily]
#!/bin/bash

if ! test -f "generated/shexml.jar"; then
    mkdir generated
    curl -L https://github.com/herminiogg/ShExML/releases/download/v0.5.3/ShExML-v0.5.3.jar -o generated/shexml.jar
fi

# generates the non-framed version
java -Dfile.encoding=UTF8 -jar generated/shexml.jar -m codemeta-shexml.shexml -f json-ld -o generated/codemeta.jsonld

# converts to the framed version
groovy ../convertToFramedJsonLD.groovy generated/codemeta.jsonld codemeta-context.jsonld generated/codemeta.json
\end{lstlisting}

Furthermore, in the context of this project, and as a real use-case demonstration, I have embedded a script generating all the CodeMeta files introduced in this paper (i.e., crosswalks, the ShExML engine and DMAOG) in a GitHub Actions workflow (see Listing \ref{lst:githubActionsConfiguration}). This will be triggered whenever a new commit is received in the GitHub repository, installing the required dependencies, running the CodeMeta generation script, validating the generated outputs, and committing the generated CodeMeta files (in case they have changes). 

The repositories for ShExML and DMAOG implement another workflow which downloads its specific CodeMeta file form the centralised CodeMeta generation workflow repository and commits the file if there has been any changes to it. This particular set-up was intended to centralise the maintenance of such a solution for various repositories and serve as a comprehensive implementation to be reused by others. Nevertheless, the whole process can be easily adapted to run directly on the targeted repository if required by the final user.

Finally, this automation ensures that any changes in the research software metadata are automatically propagated and the maintenance of the final CodeMeta file becomes easily manageable within the release process already in place for the ShExML engine and DMAOG. This set-up can be replicated by any project seeking to undertake a similar CodeMeta generation workflow to the one showcased in this paper.

\begin{lstlisting}[
 caption={GitHub Actions configuration file following the YAML syntax to configure the automatic execution of the script defined in Listing \ref {lst:scriptGenerationCodeMeta}.}, label={lst:githubActionsConfiguration}, language=yaml, frame=single,
  basicstyle=\scriptsize\ttfamily]
name: Generate CodeMeta

on: push

jobs:
  generate-codemeta:
    runs-on: ubuntu-latest

    permissions:
      contents: write

    steps:
      - uses: actions/checkout@v4
        with:
          ref: ${{ github.head_ref }}

      - name: Install Groovy
        run: sudo apt install groovy
      
      - name: Run CodeMeta generation script
        run: bash generate-all.sh

      - name: Validate generated files
        run: bash validate-all.sh
      
      - uses: stefanzweifel/git-auto-commit-action@v5
        with: 
          commit_message: Automated generation of CodeMeta files using GitHub Actions
\end{lstlisting}

\section{Results}\label{sec:results}
As they stand, the resulting mapping rules only require the modification of two lines (involving the input sources) to effectively generate a CodeMeta file for a newer software release. This entails a low-maintenance solution that can be easily integrated by developers as part of their software releasing workflows and ensures that research software aligns better with the FAIR principles, making it, at the same time, more discoverable by other researchers or platforms. In contrast with other existing solutions, the introduced method allows to map and merge an arbitrary number of heterogeneous metadata sources under a single representation while offering a flexible and easily adaptable solution due to its reliance on declarative mapping rules. Moreover, the possibility of automating it decreases developers' burden to maintain such a solution as part of their technological tool set. 

In an effort to embrace and widening the user base for CodeMeta, some Open Science platforms have already implemented support for CodeMeta within their metadata acquisition methods. This is the case of the Software Heritage Archive which harvests CodeMeta files -- amongst other metadata formats -- within existing repositories in order to construct an internal metadata registry for each record which can then be used for search purposes. Similarly, the French repository HAL allows to automatically retrieve metadata from an existing CodeMeta file when submitting new research software, streamlining the process for the user who only needs to maintain one source of truth. Zenodo do not process deposits' existing CodeMeta files yet, but it allows to export its metadata as CodeMeta and any platform leveraging its contents will be able to use the hosted CodeMeta files. Potentially, this means that large and trans-national Open Science platforms such as OpenAIRE Explore (\url{https://explore.openaire.eu/}), DataCite Commons (\url{https://commons.datacite.org/}) or the recently launched EOSC EU Node (\url{https://open-science-cloud.ec.europa.eu/}) could not only build upon their existing connections (like the aforementioned one with Zenodo) but also pull the metadata provided by the developers through CodeMeta files present in the existing repositories. Ultimately, this will allow developers to be more in control of the displayed metadata -- and its up-to-dateness -- without needing to maintain each platform's version of the same metadata through different User Interfaces (UIs) and deal with different data models.

In the context of ShExML and DMAOG, this implementation has allowed me to generate and maintain CodeMeta files for both projects with a minimal ongoing maintenance, contributing to their alignment with the FAIR attributes and promoting their discoverability and reuse. This implementation has also demonstrated how, when adapted to a development context (team composition and technical stack), its adoption by additional software packages is rather straightforward. Therefore, future research software outputs produced in this same context will be able to incorporate a CodeMeta file from the very beginning at a much lower cost. Moreover, when the implementation of CodeMeta in Open Science platforms matures, it will prevent developers from maintaining different metadata records for the same artefact, making the dissemination effort more approachable. 

\section{Future work \& conclusions}\label{sec:conclusions}
In this work, I have introduced the use of declarative mapping rules for the generation of CodeMeta files from different – and heterogeneous – extrinsic metadata providers. The showcased method allows for a flexible and rapidly adaptable workflow by which more research software projects can adopt CodeMeta at scale. In order to ease the adoption of this method, and increase its support for more types of project, the following future lines of work will be undertaken. 

It is envisaged to create more examples implementing the whole catalogue of CodeMeta crosswalks as it has been done in this work for GitHub, Maven and Zenodo. This will allow offering actionable crosswalks which demonstrate the capabilities of CodeMeta straightforwardly. Moreover, when complemented with a proper UI, these generic crosswalks should enable the generation of minimal CodeMeta files from a given reference (e.g., the GitHub repository URL), lowering down the initial learning curve, and thereby this adoption burden.

Similarly, offering more declarative mapping rules templates for CodeMeta generation, encompassing different kinds of projects with diverse implementations of technologies, will greatly benefit the rapid adoption by and adaptation to other diverse projects. This can be effectively implemented using a registry in which users can contribute with their own solutions making reuse all the more simple.

In parallel to developments on visual representations for declarative mapping rules \cite{Heyvaert16}, the automatic generation of them \cite{Chaves-Fraga22}, or the use of AI tools for all kinds of data transformations \cite{Buss25, Bedagkar25, Ghazzai24}, it is possible to offer an abstraction over them which facilitates the task of adapting or developing a set of rules from scratch by non-experts users.

Therefore, this work is a first step upon which more developments can build their efforts and showcases, by its implementation in two open-source projects, how the method can be easily adapted and is flexible enough to convey metadata from a wide variety of projects. Finally, this works aspires to streamline the generation of CodeMeta files which, in turn, strives to make research software more FAIR, increasing the reproducibility of research methods and helping to realise the Open Science precepts.

\section*{Data availability}
No external data was generated as part of this article.

\section*{Code availability}
The code presented in this paper is openly available on GitHub: \url{https://github.com/herminiogg/codemeta-shexml/}, and under the following permanent DOI: \url{https://doi.org/10.5281/zenodo.17159602}

%\bibliographystyle[super]{naturemag}
%\bibliography{references}
\printbibliography

\section*{Authors contributions}
The sole author of this paper has been responsible for developing the described methods as well as for drafting, revising and submitting the manuscript. 

\section*{Competing interests}
The author declares no competing interests.

\section*{Funding}
This work has been funded by a cascading grant entitled ``2nd open call for Route 2 support - \#1 Assessing and improving Research Software'' provided by the FAIR-IMPACT project which, in turn, has received funding from the European Commission's Horizon Europe funding programme for research and innovation programme under the Grant Agreement no. 101057344; and by the OSCARS project, which has received funding from the European Commission’s Horizon Europe Research and Innovation programme under grant agreement No. 101129751.

\end{document}